\documentclass[journal,twoside,web]{ieeecolor}
\usepackage{generic}
\usepackage{cite}
\usepackage{amsmath,amssymb,amsfonts}
\usepackage{algorithmic}
\usepackage{graphicx}
\usepackage{textcomp}
\def\BibTeX{{\rm B\kern-.05em{\sc i\kern-.025em b}\kern-.08em
    T\kern-.1667em\lower.7ex\hbox{E}\kern-.125emX}}
\usepackage{lipsum}
\usepackage{dsfont}
\usepackage{algorithm}
\usepackage{array}
\usepackage[caption=false,font=normalsize,labelfont=sf,textfont=sf]{subfig}
\usepackage{stfloats}
\usepackage{url}
\usepackage{verbatim}
\usepackage{float}
\usepackage{mathtools}
\DeclareMathOperator*{\argmax}{arg\,max}

\begin{document}
\title{PCH-EM: A solution to information loss in the photon transfer method}
\author{
Aaron J.~Hendrickson, David P.~Haefner, Stanley H.~Chan \IEEEmembership{(Senior Member, IEEE)}, Nicholas R. Shade \IEEEmembership{(Student Member, IEEE)}, and Eric R. Fossum \IEEEmembership{(Life Fellow, IEEE)}
\thanks{This work was supported in part by Naval Innovative Science \& Engineering (NISE) funding under project no.~219BAR-24-043.}
\thanks{A. J. Hendrickson is with the U.S.~Navy (NAWCAD), Patuxent River, MD 20670, USA (aaron.j.hendrickson2.civ@us.navy.mil).}
\thanks{D. P. Haefner is with the U.S.~Army (C5ISR Center), Fort Belvoir, VA 22060, USA.}
\thanks{S. H. Chan is with the School of Electrical and Computer Engineering, Purdue University, Lafayette, IN 47907, USA (stanchan@purdue.edu).}
\thanks{N. R. Shade is with the Thayer School of Engineering at Dartmouth, Dartmouth College, Hanover, NH 03755, USA (nicholas.r.shade.th@dartmouth.edu).}
\thanks{E. R. Fossum is with the Thayer School of Engineering at Dartmouth, Dartmouth College, Hanover, NH 03755, USA (eric.r.fossum@dartmouth.edu).}
}

\maketitle

\begin{abstract}
Working from a Poisson-Gaussian noise model, a multi-sample extension of the Photon Counting Histogram Expectation Maximization (PCH-EM) algorithm is derived as a general-purpose alternative to the Photon Transfer (PT) method. This algorithm is derived from the same model, requires the same experimental data, and estimates the same sensor performance parameters as the time-tested PT method, all while obtaining lower uncertainty estimates. It is shown that as read noise becomes large, multiple data samples are necessary to capture enough information about the parameters of a device under test, justifying the need for a multi-sample extension. An estimation procedure is devised consisting of initial PT characterization followed by repeated iteration of PCH-EM to demonstrate the improvement in estimate uncertainty achievable with PCH-EM; particularly in the regime of Deep Sub-Electron Read Noise (DSERN). A statistical argument based on the information theoretic concept of sufficiency is formulated to explain how PT data reduction procedures discard information contained in raw sensor data, thus explaining why the proposed algorithm is able to obtain lower uncertainty estimates of key sensor performance parameters such as read noise and conversion gain. Experimental data captured from a CMOS quanta image sensor with DSERN is then used to demonstrate the algorithm's usage and validate the underlying theory and statistical model. In support of the reproducible research effort, the code associated with this work can be obtained on the MathWorks File Exchange (Hendrickson et al., 2024).
\end{abstract}


\renewcommand{\arraystretch}{1.3} 



\section{Introduction}

\IEEEPARstart{E}{ver} since the apparent identification of photon shot noise in early Vidicon tubes employed in NASA's space missions, researchers have sought enhanced and more precise techniques for characterizing the continually advancing landscape of imaging technologies \cite{Hunten_1976,janesick_2007,janesick_2024}. In 1969, Smith \& Boyle's CCD image sensor emerged as a groundbreaking invention; a technology later made practical by Tompsett's frame transfer implementation in 1971 \cite{tompsett_1971,sequin_1973,fossum_2024}. These devices exhibited notably diminished read noise compared to Vidicon tubes, facilitating the distinct observation of shot noise \cite{janesick_2007}. During this period, Janesick \& Elliott popularized the Photon Transfer (PT) methodology for the characterization of CCDs, a technique later applied to CMOS image sensor technology and integrated into standards like the EMVA 1288 \cite{EMVA_1288_4_linear}.

At the core of the PT methodology lies the fundamental principle that raw sensor data captured over many exposure levels should be distilled into means (signal measurements) and variances (noise-squared measurements), wherein the relationships between these sample moments carry information about important sensor performance parameters such as read noise and conversion gain. From a historical viewpoint, the development of PT coincided with an era where read noise was on the order of multiple electrons. At this level of read noise, the histograms produced by a pixel provided with constant illumination are effectively normally distributed. Given that the sample mean and variance serve as the complete minimal sufficient statistic of the normal model, this PT data reduction strategy can be viewed, from a post hoc perspective, as the optimal approach at high noise; reducing the raw data to a set of summary statistics, all while preserving the information about the unknown sensor parameters contained in it (see Appendix \ref{sec:sufficient_statistics}).

In 2004, Fossum conceived the Quanta Image Sensor (QIS, pronounced `\emph{quiz}') as the next paradigm shift in digital imaging, surpassing conventional CCD and CMOS technologies \cite{DSC_book, Fossum2016_every_photon_counts}. By 2015, the first CMOS QIS was realized, demonstrating the ability to perform accurate photon number resolving by attaining Deep Sub-Electron Read Noise (DSERN) \cite{fossum_2015}. This DSERN was accomplished through a combination of reducing the capacitance of the floating diffusion sense node, thereby increasing $V/e\text{-}$ gain, and reducing source follower noise \cite{fossum_2015, fossum_2015_2, seo_2015, masoodian_2016, ma_2022}. Consequently, CMOS QISs married accurate photon number resolving, facilitated by DSERN, with the high-speed readout and radiation hardness of CMOS technology. Since the advent of CMOS QIS, DSERN has also been realized in CCD and CMOS architectures via nondestructive `Skipper' readout modes \cite{tiffenberg_2017,cervantes_2023,lapi_2024}.

\begin{figure}[htb]
    \centering
    \includegraphics[scale=1]{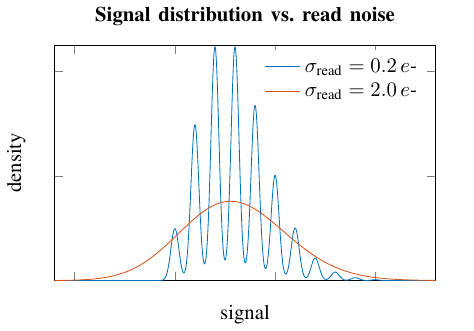}
    \caption{Pixel signal distribution for small and large read noise.}
    \label{fig:small_v_large_noise}
\end{figure}

As a result of DSERN, the histograms generated by CMOS QIS pixels exhibit markedly non-normal distributions, featuring structures previously hidden by the relatively large read noise present in traditional CCD and CMOS devices. For example, Fig.~\ref{fig:small_v_large_noise} shows the signal distribution for the same pixel at two different levels of read noise $(\sigma_\text{read})$. At large (multi-electron) noise, the distribution is effectively normal, whereas at small (deep sub-electron) noise, periodic structures appear in the distribution. The presence of these periodic structures correspond to higher information content contained in the sensor data, which cannot be fully captured by sample means and variances. Subsequently, following 2015, new characterization methods were discovered that could leverage this additional information to construct lower uncertainty estimators of the relevant sensor parameters compared to what is possible with the PT methodology \cite{starkey_2016, dutton_2016, Nakamoto_2022, hendrickson_2023_PCHEM_theory, hendrickson_comparative_study}.

Among these nascent techniques for characterization, Hendrickson \& Haefner's Photon Counting Histogram Expectation Maximization (PCH-EM) algorithm displayed potential, enabling the computation of Maximum Likelihood Estimates (MLEs) of the sensor parameters from a single data sample when the sensor being tested operates within the DSERN regime \cite{hendrickson_2023_PCHEM_theory, hendrickson_2023_PCHEM_verification}. Drawing inspiration from the PT method, PCH-EM can be generalized to integrate data from multiple samples taken at distinct exposure levels, enhancing the precision of the parameter estimates and extending the algorithm outside the DSERN regime. Remarkably, this multi-sample extension of PCH-EM utilizes the same experiment and estimates the same parameters as PT, all while delivering estimates with reduced uncertainty and superior statistical properties. Therefore, a multi-sample extension of PCH-EM can be perceived as a general purpose, reduced uncertainty, alternative to the PT method.

In this correspondence, a multi-sample generalization of the PCH-EM algorithm is derived and implemented\footnote{For a full demonstration of the implemented algorithm see the MathWorks File Exchange submission \cite{MultiSample_PCHEM_code_2024} and the discussion in Section \ref{sec:experimental_demonstration}.}. To achieve this, an overview of the Photon Counting Distribution (PCD) model, a statistical framework capable of describing data emanating from CCD, CMOS, and CMOS QIS architectures is first provided (Section \ref{sec:the_PCD_model}). Subsequently, the PT method (Section \ref{sec:the_PT_method}) and the multi-sample PCH-EM algorithm (Section \ref{sec:Multi_Sample_PCH-EM_algorithm}) is deduced from this model and an argument for the need of a multi-sample method is expounded upon (Section \ref{sec:multiple_samples}). Given that PCH-EM requires initial starting points, a statistical procedure is formulated based on initial PT estimation, subsequently refining these preliminary estimates through repeated iteration of the PCH-EM algorithm. Synthetic data from a sensor with DSERN is then subjected to parameter estimation using both the PT and PCH-EM algorithms, thereby demonstrating the enhancements achievable through PCH-EM (Section \ref{sec:monte_carlo_demonstration}). Leveraging the concept of sufficient statistics, a theoretical justification is presented to explain why PCH-EM outperforms PT, particularly within the DSERN domain, which is tied to the lossy compression PT performs when reducing raw sensor data to means and variances (Section \ref{sec:means_var_are_not_sufficient}). Finally, experimental data captured from an early CMOS QIS sensor is used to demonstrate the algorithm's usage and validate theoretical insights (Section \ref{sec:experimental_demonstration}). This work culminates with a discussion on future avenues of research and potential enhancements for the algorithm's refinement.


\section{The Photon Counting Distribution Model}
\label{sec:the_PCD_model}

The Photon Counting Distribution (PCD) describes a statistical model for data produced by CCD, CMOS, and CMOS QIS pixels in the presence of a constant signal flux. Here, a slightly altered parameterization of the PCD model is introduced, deviating from the original presentation as detailed in \cite{hendrickson_2023_PCHEM_theory}, which is commonly used in applications tied to astronomy and microscopy \cite{zhang_2007,delpretti_2008,begovic_2011,bajic_2016,fatima_2022,Barmherzig:22,hu2023poisson}.

The PCD random variable is characterized by the Poisson-Gaussian mixture model
\begin{equation}
    \label{eq:PCD_rand_var}
    X=\operatorname{round}\biggl(\frac{1}{g}\Bigl(\underbrace{\operatorname{Poisson}(H)}_{\text{electron \# $(K)$}}+\underbrace{\mathcal N(0,\sigma_{\text{read}}^2)}_{\text{read noise $(R)$}}\Bigr)+\mu\biggr).
\end{equation}
Here, $K\sim\operatorname{Poisson}(H)$ signifies the electron number, describing the number of electrons (freed by photons and/or thermal energy) associated with the gray value $X$ while $H=\mathsf EK$ denotes the quanta exposure (expected value of $K$) in units of electrons ($e\text{-}$). Additionally, $R\sim\mathcal N(0,\sigma_\text{read}^2)$ with $\sigma_\text{read}$ representing the magnitude of input-referred analog read noise in electrons ($e\text{-}$). The conversion gain is denoted as $g$ in units of electrons per digital number ($e\text{-}/\mathrm{DN}$), and $\mu$ stands for the DC offset in digital numbers ($\mathrm{DN}$). In short, the digital gray value $X$ represents a number of electrons $K$ generated in the pixel with additive read noise $R$ after application of gain, offset, and finally quantization.

The quantization described in (\ref{eq:PCD_rand_var}) adds significant complexity to the distribution of $X$. To simplify, suppose $g\leq\sigma_\text{read}$ so that the quantization error can be treated as a minor additive noise source following a Gaussian distribution $\mathcal N(0,\sigma_\text{quan}^2)$ with $\sigma_\text{quan}\approx g/\sqrt{12}$ \cite{janesick_2007}. This assumption is rather innocuous as properly designed sensors typically require conversion gain to be less than or equal to the read noise to mitigate the effects of quantization noise in their imagery \cite{janesick_2001,janesick_2007,Bohndiek_2008}. Using this simplification leads to the approximation 
\begin{equation}
\begin{aligned}
    \label{eq:PCD_rand_var_approx}
    X
    &\approx\frac{1}{g}\Bigl(\operatorname{Poisson}(H)+\mathcal N(0,\underbrace{\sigma_{\text{read}}^2+\sigma_{\text{quan}}^2}_{\sigma^2})\Bigr)+\mu\\
    &\sim\operatorname{PCD}(H,g,\mu,\sigma),
\end{aligned}
\end{equation}
where $\sigma$ is the combined read plus quantization noise in electrons ($e\text{-}$). In what follows, $\sigma$ will also be referred to as ``read noise'' and the distinction between $\sigma_\text{read}$ and $\sigma$ will largely be ignored.

Since the conditional density of $X$ given $K$ is normal, i.e., $X|K\sim\mathcal N(\mu+K/g,(\sigma/g)^2)$, it follows that the joint distribution of the complete data $(X,K)$ is given by
\begin{equation}
\label{eq:complete_data_dist}
    f_{XK}(x,k|\theta)=\frac{e^{-H}H^k}{k!}\phi(x;\mu+k/g,(\sigma/g)^2),
\end{equation}
where $\theta=(H,g,\mu,\sigma)$ denotes the PCD parameters and
\begin{equation}
    \phi(x;\alpha,\beta^2)=\frac{1}{\sqrt{2\pi}\beta}\exp\left(-\frac{(x-\alpha)^2}{2\beta^2}\right)
\end{equation}
denotes the Gaussian probability density with mean $\alpha$ and variance $\beta^2$. The PCD is then derived by integrating the complete data distribution across all possible states of the unobservable electron number $K$ yielding
\begin{equation}
\label{eq:PCD_per_pixel}
    f_X(x|\theta)=\sum_{k=0}^\infty\frac{e^{-H}H^k}{k!}\phi(x;\mu+k/g,(\sigma/g)^2).
\end{equation}
Alternatively, one may also derive an integral representation for the PCD by inverting its characteristic function $\varphi_X(t)=\mathsf Ee^{itX}$ \cite{hendrickson_comparative_study}; however, the series representation (\ref{eq:PCD_per_pixel}) is generally advisable for practical computation.


\section{Photon Transfer}
\label{sec:the_PT_method}

Photon transfer has been a well-established characterization methodology since the 1970s, utilized for estimating parameters of the PCD model from raw sensor data. The method generally involves capturing experimental data at multiple exposures, i.e., multiple $H$-values, via one of two approaches: 1) maintaining a constant source flux while varying integration time, or 2) keeping integration time constant while adjusting source flux. While there are many versions of the PT method, all of these versions share in common a data reduction strategy involving the reduction of raw data to sample means and variances followed by regression procedures to estimate the parameters. For the sake of brevity, only a constant flux version of the technique will be derived here.

Suppose $\Phi\,(e\text{-}/s)$ is the electron flux and $\tau\, (s)$ is the integration time. For this derivation the flux will be assumed to be independent of $\tau$. In this manner, $\Phi$ represents the rate at which electrons are freed so that the quanta exposure observed by the pixel is $H=\Phi\tau$. Note that if no light source is used in the experiment then $\Phi$ can be directly interpreted as the pixel's dark current. Following (\ref{eq:PCD_rand_var_approx}) the mean and variance of the pixel output can be parameterized in terms of integration time as
\begin{equation}
    (\mathsf EX)(\tau)=\Phi\tau/g+\mu
\end{equation}
and
\begin{equation}
    (\mathsf{Var}X)(\tau)=\Phi\tau/g^2+(\sigma/g)^2,
\end{equation}
respectively.

To perform the constant flux PT experiment, consider a set of $J$ integration times $\boldsymbol\tau=(\tau_1,\dots,\tau_J)$, each with a corresponding random variable
\begin{equation}
X_j\sim\operatorname{PCD}(\Phi\tau_j,g,\mu,\sigma).    
\end{equation}
From $X_j$, $N_j$ i.i.d.~observations of the pixel are drawn resulting in the random sample
\begin{equation}
    \mathbf x_j=(x_{j,1},\dots,x_{j,N_j}).
\end{equation}
Random samples captured at all $J$ integration times combined represent the multi-sample dataset $\mathbf x=(\mathbf x_1,\dots,\mathbf x_J)$. Since the integration times are known, this effectively renders the multi-sample dataset with four unknown parameters to be estimated: $\theta=(\Phi,g,\mu,\sigma)$.

Once data is captured, the estimation procedure begins by reducing the multi-sample data to means and variances resulting in
\begin{equation}
    \begin{aligned}
        \bar{\mathbf x} &=(\bar x_1,\bar x_2,\dots,\bar x_J)\\
        \hat{\mathbf x} &=(\hat x_1,\hat x_2,\dots,\hat x_J),
    \end{aligned}
\end{equation}
where
\begin{equation}
    \bar x_j=\frac{1}{N_j}\sum_{n=1}^{N_j}x_{jn}
\end{equation}
and
\begin{equation}
    \hat x_j=\frac{1}{N_j}\sum_{n=1}^{N_j}(x_{jn}-\bar x_j)^2
\end{equation}
are the sample mean and variance of $\mathbf x_j$, respectively. It is noted that throughout this work, bar-notation $(\bar{\cdot})$ is used to denote means while hat-notation $(\hat\cdot)$ denotes variances and covariances.

Following data reduction, a best fit line is estimated for the data $(\boldsymbol\tau,\bar{\mathbf x})$, yielding $\bar x(\tau)=m_1\tau+b_1\approx (\Phi/g)\tau+\mu$. Similarly, another best fit line is estimated for the data $(\boldsymbol\tau,\hat{\mathbf x})$, resulting in $\hat x(\tau)=m_2\tau+b_2\approx (\Phi/g^2)\tau+(\sigma/g)^2$. With the slopes and intercepts of these best fit lines, the PCD parameters for the pixel are estimated via
\begin{equation}
\label{eq:fluxPT_estimators}
\begin{aligned}
    \tilde\Phi &=\tilde g m_1\\
    \tilde g &=m_1/m_2\\
    \tilde\mu &=b_1\\
    \tilde\sigma &=\tilde g\sqrt{b_2}.
\end{aligned}
\end{equation}


\section{Multi-Sample PCH-EM}
\label{sec:Multi_Sample_PCH-EM_algorithm}

In the PT method, raw multi-sample data was reduced to means and variances and the linear relationships between the means and variances was exploited to characterize the pixel under test. In contrast to this approach, the PCH-EM algorithm performs Maximum Likelihood Estimation (MLE) on the same multi-sample data by reducing it to a log-likelihood function
\begin{equation}
\label{eq:sample_likelihood}
    \ell(\theta|\mathbf x)=\sum_{j=1}^J\sum_{n=1}^{N_j}\log\sum_{k=0}^\infty\frac{e^{-H_j}H_j^k}{k!}\phi(x_{jn};\mu+k/g,(\sigma/g)^2)
\end{equation}
and then determining the parameters that maximize $\ell$. The consequences of these two opposing data reduction strategies on estimator uncertainty is discussed in detail in Section \ref{sec:means_var_are_not_sufficient}.

As is common with many MLE problems, an explicit expression for the MLEs $\tilde\theta=\argmax_\theta\ell(\theta|\mathbf x)$ is intractable. Numerical methods such as the Newton-Raphson iteration may be used to numerically maximize the log-likelihood; however, poor initial guesses can lead to numerical instability when inverting the Hessian matrix so this approach can be undesirable \cite{gupta_EM_alg}.

A popular, more stable, alternative for maximizing log-likelihood functions is the Expectation Maximization (EM) algorithm, which instead of maximizing the log-likelihood directly, iteratively maximizes a related, often simpler, function to compute MLEs \cite{dempster_1977}. To see how, note that through an application of Bayes' theorem one may decompose the log-likelihood (\ref{eq:sample_likelihood}) into
\begin{equation}
    \ell(\theta|\mathbf x)=Q(\theta|\theta^{(t)})+H(\theta|\theta^{(t)}),
\end{equation}
where $Q(\theta|\theta^{(t)})=\mathsf E_{\theta^{(t)}}(\log f_{XK}(\mathbf x,\mathbf k|\theta))$ is the \emph{expected complete-data log-likelihood}. Here, the expectation is taken w.r.t. the conditional distribution $p_{K|X}(\mathbf k|\mathbf x,\theta^{(t)})$ and $\theta^{(t)}$ is a guess of the PCD parameters. Defining
\begin{equation}
    \theta^{(t+1)}=\argmax_\theta Q(\theta|\theta^{(t)})
\end{equation}
one can guarantee
\begin{equation}
    \ell(\theta^{(t+1)}|\mathbf x)-\ell(\theta^{(t)}|\mathbf x)\geq Q(\theta^{(t+1)}|\theta^{(t)})-Q(\theta^{(t)}|\theta^{(t)})
\end{equation}
showing that the updated parameter $\theta^{(t+1)}$ has a log-likelihood greater than or equal to the initial guess (in practice the increase can be substantial). By repeating the process of maximizing $Q$, each time redefining it in terms of the current parameter estimate, one can monotonically increase sample likelihood with each iteration. The benefit of doing this in the specific use case of PCH-EM is that the complete data distribution (\ref{eq:complete_data_dist}) forms an exponential family; thus, $Q$ can be maximized in closed-form resulting in a set of update equations that monotonically ``climb'' the log-likelihood function without having to evaluate derivatives or perform matrix inversions.


\subsection{Derivation of $Q$}
\label{subsec:derivation_of_Q}

To determine the form of $Q$ for multi-sample PCH-EM, call on its definition to write
\begin{multline}
    Q(\theta|\theta^{(t)})=\prod_{l=1}^J\prod_{m=1}^{N_l}\sum_{k_{lm}=0}^\infty p_{K|X}(k_{lm}|x_{lm},\theta_j^{(t)})\\
    \times\sum_{j=1}^J\sum_{n=1}^{N_j}\log f_{XK}(x_{jn},k_{jn}|\theta_j).
\end{multline}
By bringing the sums over $j$ and $n$ to the outside, all the series over $k_{lm}$ equal one, except when $l=j$ and $m=n,$ resulting in
\begin{multline}
    Q(\theta|\theta^{(t)})=\\
    \sum_{j=1}^J\sum_{n=1}^{N_j}\sum_{k=0}^\infty p_{K|X}(k|x_{jn},\theta_j^{(t)})\log f_{XK}(x_{jn},k|\theta_j).
\end{multline}
Upon substituting appropriate quantities, the final form of $Q$ is realized
\begin{multline}
    \label{eq:Q_function_general}
    Q(\theta|\theta^{(t)})=
    \sum_{j=1}^J\sum_{n=1}^{N_j}\sum_{k=0}^\infty \gamma_{jnk}^{(t)}\biggl(-H_j+k\log H_j\\
    -\log (\sigma/g)-\frac{(x_{jn}-\mu-k/g)^2}{2(\sigma/g)^2}+C\biggr),
\end{multline}
where $C$ is a constant independent of $\theta$ and
\begin{equation}
    \label{eq:gamma_nk}
    \gamma_{jnk}^{(t)}=\frac{\frac{e^{-H_j}H_j^k}{k!}\phi(x_{jn};\mu+k/g,(\sigma/g)^2)}{\sum_{\ell=0}^\infty\frac{e^{-H_j}H_j^\ell}{\ell!}\phi(x_{jn};\mu+\ell/g,(\sigma/g)^2)}\Bigg|_{\theta=\theta^{(t)}}.
\end{equation}
Here, the \emph{membership probabilities} $\gamma_{jnk}^{(t)}$ represent the probability that the electron number $K$ equals $k$ given the observed gray value $x_{jn}$ and the current parameter estimate $\theta^{(t)}$.


\subsection{Multi-Sample PCH-EM Update Equations}
\label{subsec:Multi_Sample_PCH-EM_updates}

Similar to PT, multi-sample PCH-EM data capture can be performed in either a constant flux or constant exposure mode. Combinations of these approaches can also be accommodated by specifying the form of each $H_j$ in the $Q$-function. Note that the ability to modify $Q$ to reflect the specifics of the experiment performed shows that multi-sample PCH-EM should be considered a family of algorithms and not one single set of update equations. Mimicking the previous section, a strictly constant flux version of the algorithm will be derived here. It is noted that a similar procedure can be followed to derive the constant exposure method.

Begin with the same multi-sample data captured in PT: $\mathbf x=(\mathbf x_1,\dots,\mathbf x_J)$, with $\mathbf x_j=(x_{j,1},\dots,x_{j,N_j})$ representing the $j$th sample comprised of $N_j$ i.i.d.~observations $x_{j,n}\overset{\mathrm{iid}}{\sim}\operatorname{PCD}(\Phi\tau_j,g,\mu,\sigma)$, along with the list of integration times $\boldsymbol\tau=(\tau_1,\dots,\tau_J)$ associated with each sample. Since a strictly constant flux approach is taken, $H_j\mapsto \Phi\tau_j$ for all $j\in\{1,\dots,J\}$ is substituted into the $Q$-function (\ref{eq:Q_function_general}) and $Q$ is then maximized by solving for the critical point
\begin{equation}
    \nabla_{\!\theta} Q =0
\end{equation}
with $\nabla_\theta$ denoting the gradient w.r.t.~$\theta=(\Phi,g,\mu,\sigma)$. The solution to this system, derived in Appendix \ref{sec:appendix_PCHEM_derivation}, yields the update equations
\begin{subequations}\label{eq:update_eqns}
\begin{align}
\Phi^{(t+1)} &=\bar{k}^{(t)}/\bar\tau \label{eq:Phi_update}\\
g^{(t+1)}&=\hat{k}^{(t)}/\widehat{xk}^{(t)} \label{eq:g_update}\\
\mu^{(t+1)}&=\bar{x}-\bar{k}^{(t)}/g^{(t+1)} \label{eq:mu_update}\\
\sigma^{(t+1)} &= (\hat{x}(g^{(t+1)})^2-\hat{k}^{(t)})^{1/2},\label{eq:sigma_update}
\end{align}
\end{subequations}
where
\begin{equation}
    \bar\tau=\frac{1}{N}\sum_{j=1}^JN_j\tau_j
\end{equation}
with
\begin{equation}
\label{eq:total_sample_size}
    N=\sum_{j=1}^JN_j,
\end{equation}

\begin{subequations}\label{eq:update_moments}
\begin{align}
\bar{x} &=\frac{1}{N}\sum_{j=1}^J\sum_{n=1}^{N_j}x_{jn}\label{eq:x_bar}\\
\hat{x} &=\frac{1}{N}\sum_{j=1}^J\sum_{n=1}^{N_j}(x_{jn}-\bar{x})^2\label{eq:x_hat}\\
\bar{k}^{(t)} &=\frac{1}{N}\sum_{j=1}^J\sum_{n=1}^{N_j}\sum_{k=0}^\infty\gamma_{jnk}^{(t)}k\label{eq:kj_bar_update}\\
\hat{k}^{(t)} &=\frac{1}{N}\sum_{j=1}^J\sum_{n=1}^{N_j}\sum_{k=0}^\infty\gamma_{jnk}^{(t)}(k-\bar{k}^{(t)})^2\label{eq:k_hat_update}\\
\widehat{xk}^{(t)} &=\frac{1}{N}\sum_{j=1}^J\sum_{n=1}^{N_j}\sum_{k=0}^\infty\gamma_{jnk}^{(t)}(x_{jn}-\bar{x})(k-\bar{k}^{(t)})\label{eq:xk_hat},
\end{align}
\end{subequations}
and $\gamma_{jnk}^{(t)}$ is given by (\ref{eq:gamma_nk}) after the substitution $H_j\mapsto\Phi\tau_j$.

The statistics comprising (\ref{eq:update_moments}) approximate the sufficient statistic of the complete multi-sample dataset $(\mathbf x,\mathbf k)$ from only the incomplete data $(\mathbf x)$ and the current estimate of the parameters $\theta^{(t)}=(\Phi^{(t)},g^{(t)},\mu^{(t)},\sigma^{(t)})$. Here, $\mathbf k$ are the electron numbers associated with each gray value in $\mathbf x$.  In this way,
\begin{equation}
    \label{eq:barK_approx}
    \bar k^{(t)}\approx\frac{1}{N}\sum_{j=1}^J\sum_{n=1}^{N_j}k_{jn}
\end{equation}
and
\begin{equation}
    \label{eq:hatK_approx}
    \hat k^{(t)}\approx\frac{1}{N}\sum_{j=1}^J\sum_{n=1}^{N_j}(k_{jn}-\bar k)^2
\end{equation}
are estimates of the mean and variance of the unknown electron numbers at iteration $t$ while
\begin{equation}
    \label{eq:hatXK_approx}
    \widehat{xk}^{(t)}\approx\frac{1}{N}\sum_{j=1}^J\sum_{n=1}^{N_j}(x_{jn}-\bar x)(k_{jn}-\bar k)
\end{equation}
is the estimate of the covariance between the unknown electron numbers and their corresponding known gray values at iteration $t$.

With these relationships it is straightforward to see why the update equations in (\ref{eq:update_eqns}) are of the form they assume. For example, without even placing any distributional assumptions on $K$ or $R$, assuming only that they are independent, one can write
\begin{equation}
    \mathsf{Cov}(X,K)=\mathsf{Cov}(\tfrac{K+R}{g}+\mu,K)=(\mathsf{Var}K)/g,
\end{equation}
which implies $g=\mathsf{Var}K/\mathsf{Cov}(X,K)$; matching the form of (\ref{eq:g_update}). Likewise, if one adds the additional assumption that $\mathsf{Var}R=\sigma^2$ (without necessarily assuming normality), then
\begin{equation}
    \mathsf{Var}X=(\mathsf{Var}K+\sigma^2)/g^2,
\end{equation}
which implies $\sigma=((\mathsf{Var}X)g^2-\mathsf{Var}K)^{1/2}$; matching the form of (\ref{eq:sigma_update}). Similar explanations for the forms of the remaining update equations can also be given. In fact, had the electron numbers corresponding to each $x_{jn}$ been known, the membership probabilities $\gamma_{jnk}^{(t)}$ would collapse to indicator functions $\mathds 1_{k=k_{jn}}$ (equalling one when $k=k_{jn}$ and zero otherwise), the approximations in (\ref{eq:barK_approx})-(\ref{eq:hatXK_approx}) would become equalities, and the updates in (\ref{eq:update_eqns}) would represent the exact MLEs for the complete data $(\mathbf x,\mathbf k)$; negating the need to perform any updates.

Supplied with an initial guess/estimate of the parameters $\theta^{(0)}=(\Phi^{(0)},g^{(0)},\mu^{(0)},\sigma^{(0)})$, the multi-sample PCH-EM algorithm iteratively updates (improves) the initial guess by: 1) calculating the statistics in (\ref{eq:update_moments}) (E-step) and then 2) updating the estimates with the update equations (M-step). These two steps are repeated until convergence to a fixed point is achieved, which is usually determined when the change in log-likelihood is negligible, i.e.,
\begin{equation}
    \ell(\theta^{(t+1)}|\mathbf x)-\ell(\theta^{(t)}|\mathbf x)\leq\epsilon
\end{equation}
for some appropriate choice of $\epsilon$.

For large read noise $(\sigma_\text{read}\gg 1\,e\text{-})$ the individual Gaussian components of the PCD exhibit significant overlap leading to slower convergence of the algorithm \cite{naim_2012}. Because of this, it is generally advisable for practical implementations of PCH-EM to also include a maximum allowable number of iterations that overrides the log-likelihood criteria  when it cannot be achieved with a reasonable amount of time.

After the algorithm converges, the electron numbers can be estimated by way of the membership probabilities via
\begin{equation}
    \tilde k_{jn}=\argmax_k\gamma_{jnk}^{(t)},
\end{equation}
which may find use in applications requiring photon number resolving.


\subsection{Efficient Computation via Histogram}
\label{subsec:histogram_implementation}

As the sample sizes grow, the time it takes for the statistics in (\ref{eq:update_moments}) to be computed increases due to the sums defining them involving more terms. However, because the multi-sample data is integer-valued, it is possible for PCH-EM to operate on the data histograms as opposed to the raw samples without incurring information loss (hence the use of ``histogram'' in the PCH-EM acronym). Generally, the number of unique gray values in each sample is much less than the corresponding sample size so that working from the histograms involves less computation, leading to a dramatic increase in the algorithm's execution speed.

To realize this speed improvement, $Q$ is rewritten in terms of the data histograms and the statistics in (\ref{eq:update_moments}) are subsequently re-derived. Let $(\mathbf p_j,\mathbf b_j)=((p_{j,1},b_{j,1}),\dots,(p_{j,\#_j},b_{j,\#_j}))$ denote the histogram of the $j$th sample $\mathbf x_j$. Here, $\mathbf b_j$ is the vector of unique gray values in $\mathbf x_j$ and $\mathbf p_j$ are sample probabilities (counts in each bin $b_{jn}$ normalized by $N_j$). Using this convention, $\#_j$ represents the number of unique gray values in the $j$th sample. The histograms subsequently allow $Q$ to be equivalently written as
\begin{multline}
     \label{eq:Q_fun_hist}
    Q(\theta|\theta^{(t)})=
    \sum_{j=1}^JN_j\sum_{n=1}^{\#_j}p_{jn}\sum_{k=0}^\infty \gamma_{jnk}^{(t)}\biggl(-H_j+k\log H_j\\
    -\log (\sigma/g)-\frac{(b_{jn}-\mu-k/g)^2}{2(\sigma/g)^2}+C\biggr),
\end{multline}
where $\gamma_{jnk}^{(t)}$ is given by (\ref{eq:gamma_nk}) with the substitutions $H_j\mapsto\Phi\tau_j$ and $x_{jn}\mapsto b_{jn}$.

Maximizing this altered form of $Q$ yields the following statistics, which may be substituted into the PCH-EM update equations (\ref{eq:update_eqns}) to improve algorithm speed:
\begin{equation}
\label{eq:bar_tau}
    \bar\tau=\sum_{j=1}^Jw_j\tau_j
\end{equation}
with
\begin{equation}
\label{eq:w_j+formula}
w_j =\frac{N_j}{\sum_{m=1}^JN_m}
\end{equation}
and
\begin{subequations}\label{eq:update_moments2}
\begin{align}
\bar{x} &=\sum_{j=1}^Jw_j\sum_{n=1}^{\#_j}p_{jn}b_{jn}\label{eq:x_bar}\\
\hat{x} &=\sum_{j=1}^Jw_j\sum_{n=1}^{\#_j}p_{jn}(b_{jn}-\bar{x})^2\label{eq:x_hat}\\
\bar{k}^{(t)} &=\sum_{j=1}^Jw_j\sum_{n=1}^{\#_j}p_{jn}\sum_{k=0}^\infty\gamma_{jnk}^{(t)}k\label{eq:kj_bar_update}\\
\hat{k}^{(t)} &=\sum_{j=1}^Jw_j\sum_{n=1}^{\#_j}p_{jn}\sum_{k=0}^\infty\gamma_{jnk}^{(t)}(k-\bar{k}^{(t)})^2\label{eq:k_hat_update}\\
\widehat{xk}^{(t)} &=\sum_{j=1}^Jw_j\sum_{n=1}^{\#_j}p_{jn}(b_{jn}-\bar{x})\sum_{k=0}^\infty\gamma_{jnk}^{(t)}(k-\bar{k}^{(t)})\label{eq:xk_hat}.
\end{align}
\end{subequations}


\section{Why adopt a multi-sample approach?}
\label{sec:multiple_samples}

Until now, constant flux PT and multi-sample PCH-EM methods have been developed without a clear reason for employing a multi-sample approach. It is important to note that when a pixel exhibits DSERN, the raw data it produces contains sufficient information to permit estimating the PCD parameters from a single sample. This assertion is supported by the fact that reasonable parameter estimates can be obtained simply by visually examining the histogram generated by a DSERN pixel. However, as the read noise increases, the information content of the data decreases, and the PCD approaches a normal distribution. In this high noise regime, the minimal sufficient statistic of the multi-sample data consists of a sample mean and variance for each sample ($J$ means and $J$ variances). For a single sample approach ($J=1$), the data collapses to a two-dimensional statistic; however, since the constant flux model has four parameters ($\theta=(\Phi,g,\mu,\sigma)$), a single sample does not provide enough information about the parameters for reliable estimation in high noise conditions. This limitation applies not only to single-sample PCH-EM but also to other single sample methods such as the Photon Counting Histogram (PCH) method, Valley Peak Modulation (VPM) method, Peak Separation and Width (PSW) method, and FFT method \cite{starkey_2016,dutton_2016,hendrickson_comparative_study}.

By adding an additional sample captured at a different integration time, the minimal sufficient statistic in high noise conditions becomes four-dimensional, matching the dimension of the unknown parameter vector. This implies enough information about the parameters has been captured to estimate them. In conclusion, the single-sample PCH-EM procedure tends toward an underspecified model in high noise conditions, while the multi-sample PCH-EM procedure converges toward a well-defined/over-determined model. Therefore, to ensure effectiveness both within and beyond the DSERN regime, a multi-sample data approach is necessary. This is particularly crucial as modern CMOS QISs exhibit a distribution of read noise across all pixels in the sensor array. While most pixels exhibit DSERN, there is typically a small subset of pixels with read noise outside the DSERN regime. By adopting a multi-sample approach, the algorithm becomes robust to higher read noise and can accurately characterize all pixels.

Combining multiple samples into a single estimation procedure not only enhances the algorithm's resilience against high read noise but also facilitates obtaining more accurate parameter estimates. For simplicity, suppose the sample sizes are large enough, so that the constant flux multi-sample PCH-EM estimates, denoted as $\tilde\theta=(\tilde\Phi,\tilde g,\tilde\mu,\tilde\sigma)^T$, follow a normal distribution $\tilde\theta\sim\mathcal N(\theta,\Sigma)$, with covariance equivalent to the inverse Fisher information, $\Sigma=\mathsf E(\tilde\theta-\theta)(\tilde\theta-\theta)^T=I_\mathbf{X}^{-1}(\theta)$. In situations with low read noise, the Fisher information can be approximated by the complete data information, leading to the generalized variance approximation
\begin{equation}
    \det\Sigma\sim(\det I_{\mathbf X\mathbf K}(\theta))^{-1}=\frac{\Phi\sigma^6/2}{N^4\,\bar\tau\,\mathsf{Var}\mathcal K}.
\end{equation}
Here, $N$ represents the total sample size (\ref{eq:total_sample_size}), $\bar\tau$ is the average integration time given in (\ref{eq:bar_tau}), and
\begin{equation}
    \mathsf{Var}\mathcal K=\Phi(\bar\tau+\Phi\hat\tau)
\end{equation}
where
\begin{equation}
    \hat\tau=\sum_{j=1}^Jw_j(\tau_j-\bar\tau)^2
\end{equation}
is the variance of the integration times with $w_j$ as defined in (\ref{eq:w_j+formula}).

\begin{figure}[htb]
    \centering
    \includegraphics[scale=1]{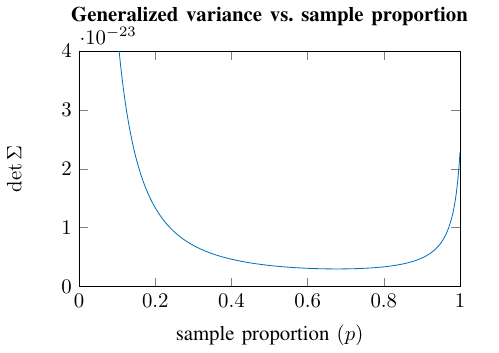}
    \caption{Two-sample generalized variance versus sample proportion $p$.}
    \label{fig:generalized_variance}
\end{figure}

Now, consider a two-sample procedure where the total sample size $N$ is fixed, so that $N_1=(1-p)N$ and $N_2=pN$ for some $p\in[0,1]$. Fig.~\ref{fig:generalized_variance} plots the generalized variance approximation as a function of $p$ for $\Phi=10\,e\text{-}/s$, $\sigma=0.15\,e\text{-}$, $N=10^4$, $\tau_1=0.1\,s$, and $\tau_2=5\,s$. At $p=0$, the generalized variance corresponds to that of a single-sample method with integration time $\tau_1$. Similarly, at $p=1$, it corresponds to that of a single-sample method with integration time $\tau_2$. The plot reveals that the generalized variance is minimized when employing a two-sample approach with $p\approx 0.7$, demonstrating that utilizing two or more samples can lead to lower variance compared to a one-sample approach, even with the same total number of observations. This phenomenon can be attributed to ``information diversification.'' The amount of information about each parameter present in the raw sensor data dynamically varies with quanta exposure $H=\Phi\tau$. For instance, with very small $H$ (i.e., small $\tau$), the data contains abundant information about $\mu$, while with large $H$ (i.e., large $\tau$), it is rich in information about $g$. By combining multiple samples with varying $H$, the collective information in the multi-sample data is diversified across all parameters, generally leading to improved overall uncertainty in the parameter estimates.


\section{Monte Carlo Demonstration}
\label{sec:monte_carlo_demonstration}

With constant flux versions of the PT method and PCH-EM algorithm derived, Monte Carlo experiments were performed to demonstrate the performance increases achievable with PCH-EM. For the experiment, a two-sample $(J=2)$ data capture was adopted and a pixel with the parameters outlined in Table \ref{tab:simulation_parameters} was specified. A total of sixty-four linearly space values of the read noise on the interval $[0.05,1]$ were chosen to help observe the effect read noise has on estimate uncertainty and the ability of PCH-EM to improve the PT estimates.

\begin{table}[htb]
\begin{center}
\caption{Simulation parameters.}
\label{tab:simulation_parameters}
\begin{tabular}{| c | c | c |}
\hline
& value & unit\\
\hline
\hline
$\Phi$ & $5$ & $e\text{-}/s$\\
\hline
$g$ & $\sigma_\text{read}/4$ & $e\text{-}/\mathrm{DN}$\\
\hline
$\mu$ & $100$ & $\mathrm{DN}$\\
\hline
$\sigma_\text{read}$ & $\operatorname{linspace}(0.05,1,64)$ & $e\text{-}$\\
\hline
$\tau_1$ & $0.01$ & $s$\\
\hline
$\tau_2$ & $1$ & $s$\\
\hline
\end{tabular}
\end{center}
\end{table}

\begin{figure}[htb]
    \centering
    \includegraphics[scale=1]{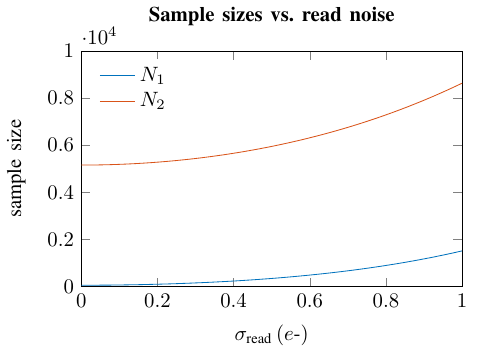}
    \caption{Samples sizes versus read noise.}
    \label{fig:sample_sizes}
\end{figure}

An important consideration in the experimental design is the sample sizes.  As read noise increases, the uncertainty in the estimates of both methods will also increase \cite{beecken_96,Hendrickson_17}. To keep the uncertainty of the estimates from growing too large, the sample sizes were determined via \cite{hendrickson_22}
\begin{equation}
    \begin{aligned}
        N_1 &= \operatorname{ceil}\left(\frac{2\zeta(1+\zeta)}{\delta^2(1-\zeta)^2}+1\right)\\
        N_2 &= \operatorname{ceil}\left(\frac{2(1+\zeta)}{\delta^2(1-\zeta)^2}+5\right),
    \end{aligned}
\end{equation}
where $\delta=0.02$ and
\begin{equation}
    \zeta=\frac{\Phi\tau_1+\sigma_\text{read}^2}{\Phi\tau_2+\sigma_\text{read}^2}.
\end{equation}
As seen in Fig.~\ref{fig:sample_sizes}, these sample sizes will increase with increasing read noise and subsequently force the relative uncertainty of the PT conversion gain estimate in (\ref{eq:fluxPT_estimators}) to be approximately equal to $\delta$ ($2\%$ error), independent of the parameter values.

The experiment was performed by estimating the parameter $\theta=(\Phi,g,\mu,\sigma)$ using PT a total of $N_\mathrm{trials}=1000$ times at each of the sixty four parameter values and then feeding these estimate into PCH-EM for refinement. To measure each method's estimation error, a Relative Root Mean Squared Error (RRMSE) metric was adopted
\begin{equation}
\begin{aligned}
    \operatorname{RRMSE}(\tilde\theta_i)
    &=\left(\frac{1}{N_\mathrm{trials}}\sum_{n=1}^{N_\mathrm{trials}}\left(\frac{\tilde\theta_{i,n}-\theta_i}{\theta_i}\right)^2\right)^{1/2}\\
    &\approx\mathsf E(\operatorname{RRMSE}(\tilde\theta_i))\\
    &=\frac{\sqrt{\mathsf{MSE}(\tilde\theta_i)}}{|\theta_i|},
\end{aligned}    
\end{equation}
where $\theta_i$ is replaced with one of $\Phi$, $g$, $\mu$, or $\sigma$, $\tilde\theta_{i,n}$ is the $n$th estimate of $\theta_i$ from either PT or PCH-EM, and $\mathsf{MSE}(\tilde\theta_i)=\mathsf E(\tilde\theta_i-\theta_i)^2$ is the exact value of the estimator Mean Squared Error (MSE).

If the sample sizes are sufficiently large, then the PCH-EM estimates should be nearly unbiased so that RRMSE can further be approximated by the relative standard deviation
\begin{equation}
    \operatorname{RRMSE}(\tilde\theta_i)\approx\frac{\sqrt{\mathsf{Var}(\tilde\theta_i)}}{|\theta_i|}.
\end{equation}
Using the Fisher information of the multi-sample data
\begin{equation}
    I_\mathbf{X}(\theta)=-\mathsf E(\nabla_\theta\nabla_\theta^T\ell(\theta|\mathbf X)),
\end{equation}
the Cram\'{e}r-Rao Lower Bound (CRLB) on the relative standard deviation for each parameter estimate can be calculated from diagonal elements of $I_\mathbf{X}^{-1}$ via
\begin{equation}
    \frac{\sqrt{(I_\mathbf{X}^{-1}(\theta))_{ii}}}{|\theta_i|}\leq\frac{\sqrt{\mathsf{Var}(\tilde\theta_i)}}{|\theta_i|}.
\end{equation}
Consequently, to compliment the simulated RRMSE data for each parameter, the CRLB was computed using a Monte Carlo method for estimating the Fisher information at each of the sixty-four parameter values.

\begin{figure}[p]
    \centering
    \includegraphics[scale=1]{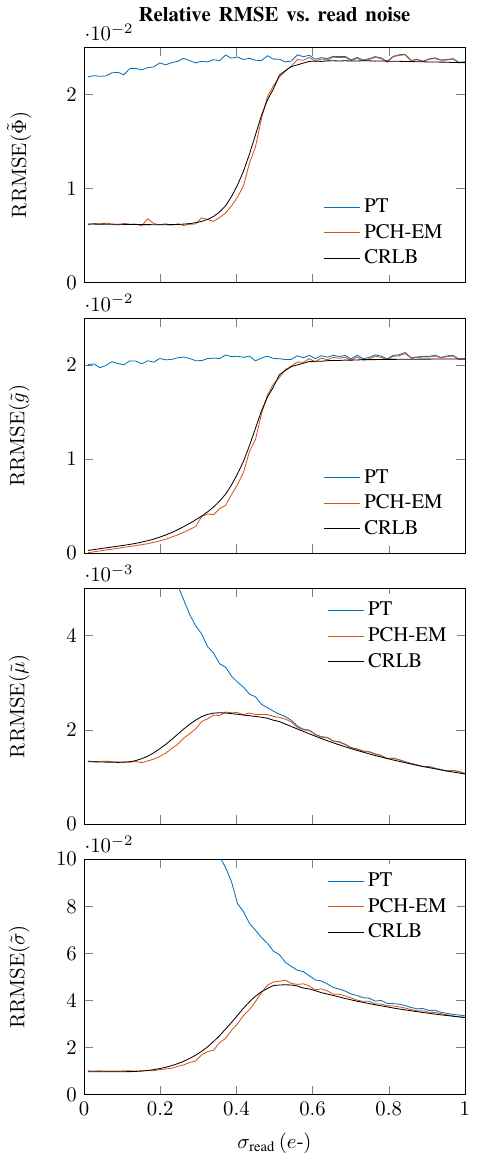}
    \caption{RRMSE vs.~read noise for parameter estimates computed using constant flux implementation of PT and PCH-EM. RRMSE curves for PT $\tilde\mu$ and $\tilde\sigma$ grow large near $\sigma_\text{read}=0$ and were clipped from the plot window.}
    \label{fig:RMSE_plots}
\end{figure}

Fig.~\ref{fig:RMSE_plots} shows the CRLB and RRMSE curves for each method and each parameter as a function of read noise. Upon inspection, the specified sample sizes did an excellent job of holding the PT conversion gain estimate's relatively uncertainty close to the specified value of $\delta=0.02$. Far more interesting is the behavior of the PCH-EM RRMSE curves with respect to those of PT. First, the relative uncertainty of the PCH-EM estimates are less than those of PT at all read noise values, particularly in the DSERN regime. For example, at $\sigma_\text{read}=0.214\, e\text{-}$ the RRMSE of the PT estimate for $\Phi$ was reduced via PCH-EM by $73\%$, while the RRMSE of the PT estimates for $g$, $\mu$, and $\sigma$, were reduced by $93\%$, $74\%$, and $96\%$, respectively. Indeed, the PCH-EM curves are nearly identical to the CRLB curves indicating that for the specified parameters, PCH-EM is producing the best possible estimates. Second, in all four parameters, the PCH-EM RRMSE curves approached those of PT, from below, asymptotically, as the read noise increases. Further explanation of this is provided in the next section. This characteristic of PCH-EM can thus be viewed as an overall improvement over the PT method since it achieves less uncertainty than PT everywhere and is asymptotically equivalent to PT at large read noise.


\section{Performance differences between PT and PCH-EM}
\label{sec:means_var_are_not_sufficient}

In the preceding section it was shown that the PCH-EM algorithm was able to improve PT estimates in terms of MSE and the magnitude of this improvement vanished with increasing read noise. The source of this phenomenon can be largely traced to the way in which both methods reduce the raw multi-sample data prior to estimation.

In PCH-EM, the estimation procedure interacts with the raw multi-sample data through a log-likelihood function, that is, the multi-sample data is reduced via
\begin{equation}
\label{eq:PCHEM_data_redux}
    \mathbf x\mapsto\ell(\theta|\mathbf x),
\end{equation}
where $\ell$ denotes the log-likelihood in (\ref{eq:sample_likelihood}). Established statistical theory states that the data reduction in (\ref{eq:PCHEM_data_redux}), when combined with the MLE equivalence relation specifying proportional likelihoods as equivalent, induces a minimal sufficient partition on the sample space \cite[$\S$ 6.6.3]{casella_2002}. In other words, the likelihood function, when paired with this equivalence relation, encodes all of the information in the data, about the unknown parameters $\theta$, as efficiently as possible without losing any of the information. This minimally sufficient encoding subsequently endows the PCH-EM estimates with particularly nice statistical properties, i.e., asymptotic normality, unbiasedness, and efficiency, ensuring that PCH-EM, at the very least, produces the best possible estimates at large sample sizes.

On the other hand, PT interacts with the multi-sample data through sample means and variances via the reduction
\begin{equation}
\label{eq:PT_data_redux}
\mathbf x\mapsto\!\!\!
\begin{array}{l}
(\bar x_1,\dots,\bar x_J)\\
(\hat x_1,\dots,\hat x_J)
\end{array}
.
\end{equation}
A test proposed by Pinelis (2023) can be used to show that the PCD does not form an exponential family \cite{Pinelis_2023}. This result, combined with the Pitman--Koopman--Darmois theorem, establishes that the reduction (\ref{eq:PT_data_redux}) cannot be sufficient, let alone minimal sufficient. As such, the data reduction (\ref{eq:PT_data_redux}) incurs a loss of information about the parameters and this information loss leads to higher uncertainty estimates in the PT method. In fact, because the PT data reduction is not sufficient, the Rao-Blackwell theorem informs us that the PT method estimators can be improved in terms of MSE by conditioning them on sufficient statistics, namely \cite[Thm.~7.3.17]{casella_2002},
\begin{equation}
    \mathsf{MSE}(\mathsf E(\tilde\theta_{\mathrm{PT}}|T(\mathbf x)))\leq \mathsf{MSE}(\tilde\theta_{\mathrm{PT}}),
\end{equation}
where $\tilde\theta_{\mathrm{PT}}$ is any one of the PT parameter estimators, $\mathsf E(\tilde\theta_\mathrm{PT}|T(\mathbf x))$ is the Rao-Blackwellised variant of $\tilde\theta_\mathrm{PT}$, and $T$ is a sufficient statistic. However, as a consequence of Neyman factorization, PCH-EM estimates can be written as functions of any sufficient statistic, including minimal sufficient statistics; hence, no such improvement in terms of MSE can be achieved via Rao-Blackwellization.

The arguments laid out establish that one should expect PCH-EM estimates to outperform PT estimates in terms of MSE but it does not explain the convergence of the MSE in both methods as read noise increases. To that end, note that as read noise increases, the PCD converges to a normal distribution of equivalent mean and variance.  This can be demonstrated by assigning $X\sim\operatorname{PCD}(H,g,\mu,\sigma)$ and showing that the characteristic function of the standardized random variable
\begin{equation}
    Z=\frac{X-\mathsf EX}{\sqrt{\mathsf{Var}X}}
\end{equation}
converges to the characteristic function of the standard normal distribution in the limit $\sigma\to\infty$. Since the sample mean and variance constitute the complete minimal sufficient statistic of the normal model, the information loss incurred in the PT data reduction (\ref{eq:PT_data_redux}) vanishes at large read noise. Consequently, at large read noise both methods optimally encode the information contained in the raw data without loss and thus produce estimates with nearly identical MSE.

\begin{figure}[htb]
    \centering
    \includegraphics[scale=1]{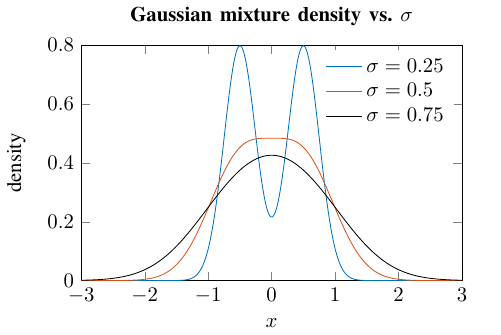}
    \caption{Gaussian mixture showing loss of peaks at $\sigma=0.5$.}
    \label{fig:gaussian_mixture}
\end{figure}

What constitutes large enough read noise for the methods to converge can be observed from Fig.~\ref{fig:RMSE_plots} as PCH-EM provides little improvement to PT above a threshold of approximately $\sigma_\text{read}=0.5\,e\text{-}$. This threshold corresponds to the upper bound of the DSERN regime as originally defined in \cite{fossum_2015} where the peaks (local maxima) in the PCD begin to vanish. As a simplified demonstration of this phenomena, consider the following two-component Gaussian mixture which mimics the properties of the PCD:
\begin{equation}
    f(x)=\frac{1}{2}\phi(x;-1/2,\sigma^2)+\frac{1}{2}\phi(x;1/2,\sigma^2).
\end{equation}
As seen in Fig.~\ref{fig:gaussian_mixture}, for $\sigma<0.5$ two peaks in this density function are observed at $x=-1/2$ and $x=1/2$ with a corresponding local minima at $x=0$. The existence of the local minima corresponds to the peaks being resolved. However, as $\sigma$ increases, the local minima increases until transitioning to a global maxima precisely when $\sigma=0.5$.

From a historical perspective, the fact that PCH-EM only begins to gain significant advantage over PT in the DSERN regime gives context to why such a method is not already in widespread use for sensor characterization.  Had PCH-EM been developed to characterize early CCDs in the 1970s, it would have exhibited little to no improvement over the much simpler and computationally friendly PT method, rendering PT the preferred technique. With the advent of DSERN technology, PCH-EM can now deliver superior estimates over PT; thus, establishing its utility.


\section{Experimental Demonstration}
\label{sec:experimental_demonstration}

\begin{figure}[htb]
    \centering
    \includegraphics[scale=0.042]{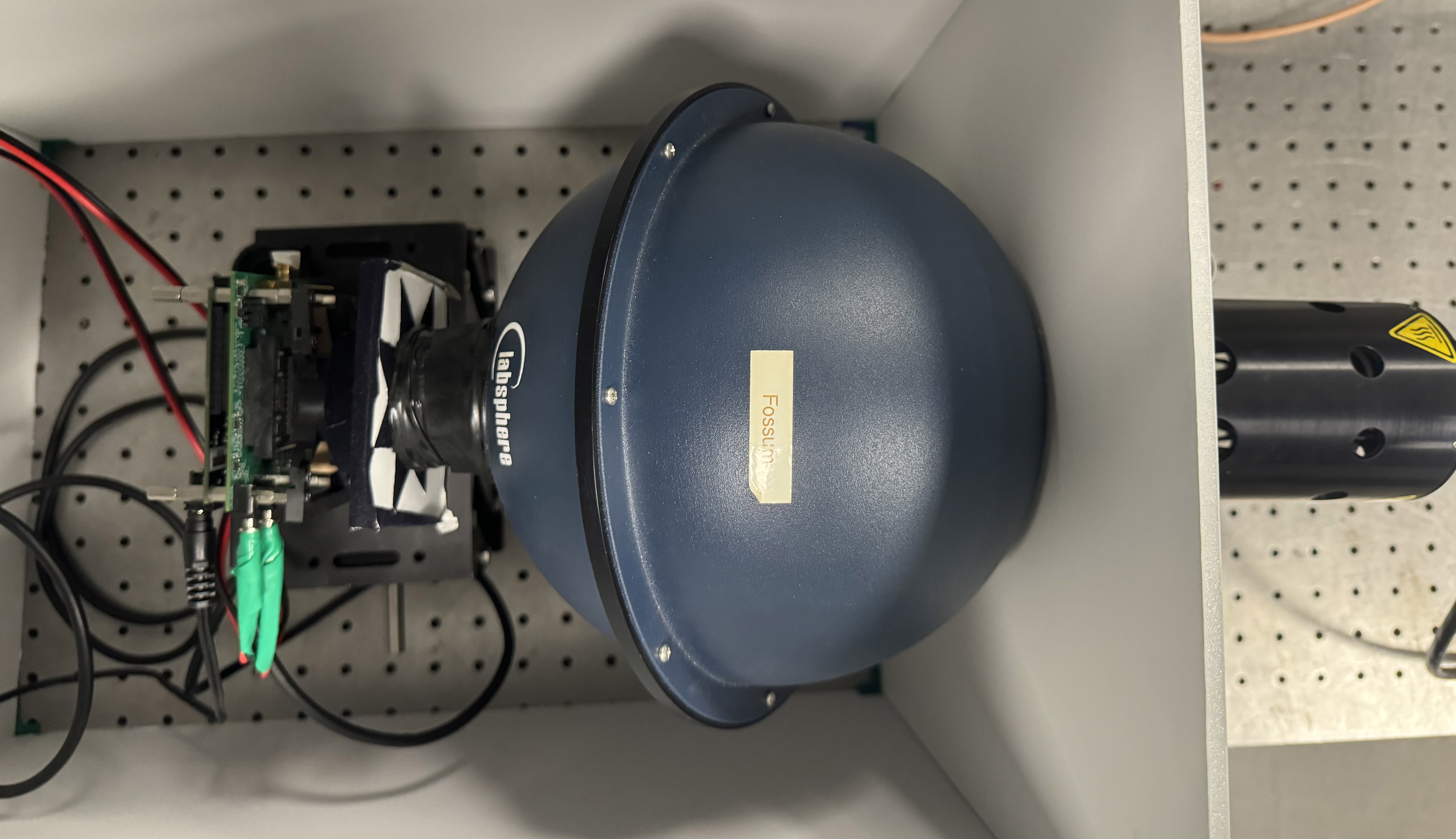}
    \caption{Cleveland camera experiment setup.}
    \label{fig:experimental_setup_photo}
\end{figure}

Experimental constant flux data was captured with the Gigajot Technology Inc.~(GJ00111) ``Cleveland'' camera.  The constant source of electron flux came from a combination of the sensor's internal dark current and impinging photons from an incandescent light source. Fig.~\ref{fig:experimental_setup_photo} shows the experimental setup consisting of the incandescent source coupled with an integrating sphere to illuminate the sensor with a stable, flat field.
For the experiment, a three-sample approach was adopted with integration times and sample sizes given in Table \ref{tab:experimental_parameters}.
\begin{table}[htb]
\begin{center}
\caption{Experimental parameters.}
\label{tab:experimental_parameters}
\begin{tabular}{| c | c | c |}
\hline
 & int. time $(\tau_j)$ & sample size $(N_j)$\\
\hline
\hline
sample 1 & $0.876\,\mathrm{ms}$ & $8500$\\
\hline
sample 2 & $3.3\,\mathrm{ms}$ & $14000$\\
\hline
sample 3 & $6.6\,\mathrm{ms}$ & $20000$\\
\hline
\end{tabular}
\end{center}
\end{table}

Experimental data was processed in Matlab using the code uploaded to the MathWorks File Exchange (FEX) \cite{MultiSample_PCHEM_code_2024}. It is noted that the code available on the FEX provides a full demonstration of constant flux PCH-EM by generating synthetic data. Here, the code was modified by removing the synthetic data generation, instead accepting raw experimental data from the Cleveland camera experiment. Minimal effort was put into optimizing the efficiency of the code with the intent to make it more readable and track with the equations laid out in this work. To that end, the code is commented with corresponding equation numbers, when feasible, to help the user follow along with this paper.

The code (version 1.1.1) follows a step-by-step procedure in the main file \texttt{demo.m} to processes the raw multi-sample data by first organizing it into the data structure \texttt{Struct} (\texttt{fluxData.m}) and then passing this structure to various steps in data processing pipeline. In \texttt{fluxData.m}, the data for each sample is reduced to a histogram and these histograms, along with the raw samples, are used throughout the processing pipeline. The data processing pipeline following \texttt{fluxData.m} consists of 1) performing initial starting point estimation (\texttt{fluxStart.m}), 2) performing PCH-EM estimation seeded with the starting points (\texttt{fluxPCHEM.m}), and 3) computing the covariance matrix for the final PCH-EM estimates (\texttt{fluxCovariance.m}).

The function \texttt{fluxStart.m} uses a hybrid approach to starting point estimation by estimating the starting points with the FFT method (\texttt{fluxFFT.m}), described in \cite{hendrickson_2023_PCHEM_theory,hendrickson_comparative_study}, if only a single sample is captured or the constant flux PT method (\texttt{fluxPT.m}) if two or more samples are captured. By using this hybrid approach, the code works for any numbers of samples. Likewise, \texttt{fluxCovariance.m} estimates the PCH-EM covariance matrix by computing the observed information matrix
\begin{equation}
    J(\tilde\theta)=-\nabla_\theta\nabla_\theta^T\ell(\theta|\mathbf x)\Big|_{\theta=\tilde\theta},
\end{equation}
where $\tilde\theta$ is the PCH-EM estimate of the parameters.  In this way, $J^{-1}$ becomes an estimate for the PCH-EM estimate covariance matrix and the diagonal elements of $J^{-1}$ represent standard error estimates for each PCH-EM parameter estimate. Additional functions are also supplied in the FEX submission to display the final estimates and provide plots pertaining to algorithm convergence and data visualization.

\begin{figure}[htb]
    \centering
    \includegraphics[scale=1]{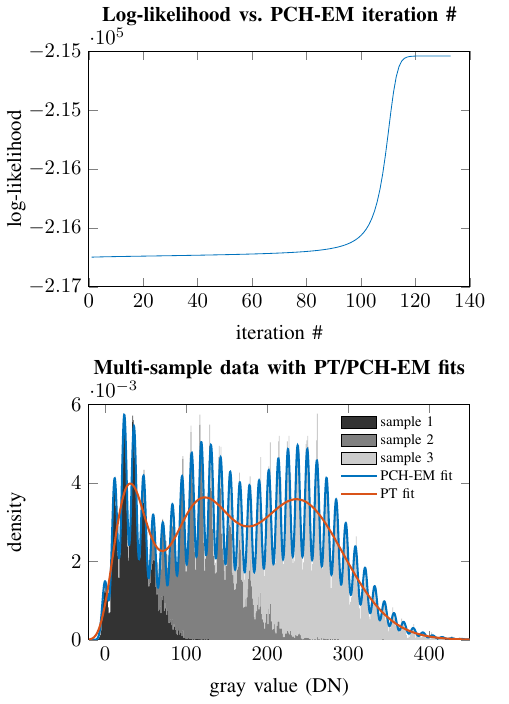}
    \caption{Log-likelihood versus PCH-EM iteration number (top) and experimental multi-sample data with PT/PCH-EM fits (bottom) for the small noise pixel in Table \ref{tab:low_noise_pixe_estimates}.}
    \label{fig:small_noise_pixel}
\end{figure}

\begin{figure}[htb]
    \centering
    \includegraphics[scale=1]{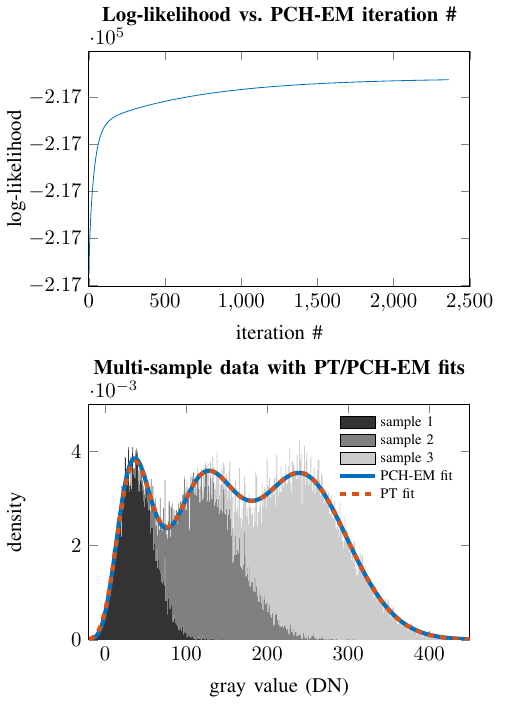}
    \caption{Log-likelihood versus PCH-EM iteration number (top) and experimental multi-sample data with PT/PCH-EM fits (bottom) for the larger noise pixel in Table \ref{tab:large_noise_pixe_estimates}.}
    \label{fig:large_noise_pixel}
\end{figure}

Tables \ref{tab:low_noise_pixe_estimates}-\ref{tab:large_noise_pixe_estimates} present the characterization results for two of the QIS pixels tested: one with small read noise and another with larger read noise. Comparing the standard errors in both tables not only reveals the impact read noise has on estimate uncertainty but also the level of precision obtainable at low read noise.  For example, the PCH-EM standard error estimate for $\tilde g$ in Table \ref{tab:low_noise_pixe_estimates} indicates approximately ten decimal places of precision. This error estimate was independently verified through Monte Carlo experiments; verifying that the observed information matrix gave accurate uncertainty estimates. Corresponding to these tables are Fig.~\ref{fig:small_noise_pixel}-\ref{fig:large_noise_pixel} showing the log-likelihood obtained at each iteration of PCH-EM (top) and quality of fit obtained with PT and PCH-EM (bottom). To visualize the quality of fit, the multi-sample data for each pixel was turned into a ``stacked'' histogram with gray shade indicating the frequency in each bin obtained from each sample. This stacked histogram was then overlaid with the estimated multi-sample PCD
\begin{equation}
    f(x|\tilde\theta)=\sum_{j=1}^Jw_jf_X(x|\tilde\theta_j),
\end{equation}
where $f_X$ is the PCD in (\ref{eq:PCD_per_pixel}), $w_j$ are the sample weights (proportion of total sample size budget apportioned to sample $j$), and $\tilde\theta_j=(\tilde\Phi\tau_j,\tilde g,\tilde\mu,\tilde\sigma)$ are the PT or PCH-EM parameter estimates.

\begin{table}[htb]
\begin{center}
\caption{Small noise pixel characterization (Fig.~\ref{fig:small_noise_pixel}).}
\label{tab:low_noise_pixe_estimates}
\begin{tabular}{| c | c | c | c |}
\hline
 & PT & PCH-EM & std. error (PCH-EM)\\
\hline
\hline
$\tilde\Phi\,(e\text{-}/s)$ & $3175$ & $3154$ & $\pm 17$\\
\hline
$\tilde g\,(e\text{-}/\mathrm{DN})$ & $0.0862859140$ & $0.0841397938$ & $\pm 7.3\times 10^{-10}$\\
\hline
$\tilde\mu\,(\mathrm{DN})$ & $2.955$ & $-0.070$ & $\pm 3.7\times 10^{-3}$\\
\hline
$\tilde\sigma\,(e\text{-})$ & $0.667807$ & $0.286082$ & $\pm 2.1\times 10^{-6}$\\
\hline
\end{tabular}
\end{center}
\end{table}

\begin{table}[htb]
\begin{center}
\caption{Larger noise pixel characterization (Fig.~\ref{fig:large_noise_pixel}).}
\label{tab:large_noise_pixe_estimates}
\begin{tabular}{| c | c | c | c |}
\hline
 & PT & PCH-EM & std. error (PCH-EM)\\
\hline
\hline
$\tilde\Phi\,(e\text{-}/s)$ & $3117$ & $3074$ & $\pm 820$\\
\hline
$\tilde g\,(e\text{-}/\mathrm{DN})$ & $0.0852220$ & $0.0840197$ & $\pm 5.6\times 10^{-7}$\\
\hline
$\tilde\mu\,(\mathrm{DN})$ & $8.18$ & $8.16$ & $\pm 7.1\times 10^{-2}$\\
\hline
$\tilde\sigma\,(e\text{-})$ & $0.909$ & $0.792$ & $\pm 1.0\times 10^{-3}$\\
\hline
\end{tabular}
\end{center}
\end{table}
			
In reference to the data in Fig.~\ref{fig:small_noise_pixel}, one can observe that the algorithm monotonically increased log-likelihood (improved the PT estimates) with each iteration and the fit obtained with PT (red curve) was substantially improved with PCH-EM (blue curve). Likewise, now referencing the data in Fig.~\ref{fig:large_noise_pixel}, one can observe that the noise is large enough to yield approximately normally distributed data and that the PCH-EM fit is nearly identical to the PT fit as is expected according to Section \ref{sec:means_var_are_not_sufficient}. As was pointed out in Section \ref{subsec:Multi_Sample_PCH-EM_updates}, at larger noise, the peaks in the PCD exhibit more overlap leading to a slowing down of the algorithm convergence as indicated by the larger number of iterations needed for PCH-EM to converge in the large noise pixel data.  Despite this slower convergence, the algorithm was still able to monotonically increase log-likelihood at each iteration. It is noted that only a small change in log-likelihood for this pixel was achieved with PCH-EM and that the $y$-axis tick labels in Fig.~\ref{fig:large_noise_pixel} (top) were unable to capture this small change with only two decimal places of precision. The ability of multi-sample PCH-EM to handle both small and large noise pixels validates its usage as a general characterization algorithm not limited to only pixels with DSERN.


\section{Discussion}
\label{sec:discussion}

Camera companies and camera consumers alike rely on characterization techniques to troubleshoot, optimize, and calibrate image sensors. As sensor technology continues to evolve, new characterization methods are developed to measure key performance parameters and inform decision making. In this correspondence, the next step in this evolution was taken by developing and implementing a multi-sample PCH-EM algorithm, which shows dramatic improvement in estimate uncertainty over the Photon Transfer (PT) method, particularly in the regime of Deep Sub-Electron Read Noise (DSERN). It was shown that the source of this uncertainty advantage is tied to the lossy compression PT performs on the raw sensor data by reducing it to sample means and variances. Since PCH-EM does not perform such a lossy compression, it is able to utilize all of the available information in the raw data, leading to lower estimate uncertainty. As read noise exceeds the DSERN threshold, the information loss incurred by PT vanishes and both methods show nearly identical estimate uncertainty; thus PCH-EM can be viewed as a general improvement over the PT method.

This maximum likelihood approach to sensor characterization, implemented via Expectation Maximization (EM), is in actuality a general approach not limited to a single algorithm or single sensor technology. By constructing various $Q$-functions reflecting an assumed sensor model and experimental parameters, a large number of PCH-EM algorithms can be derived. Here, the process of $Q$-function construction and update equation derivation was demonstrated for a single pixel, constant flux, approach to PCH-EM using multiples samples. A complementary constant flux implementation of PT was also derived to provide initial starting points to seed the algorithm. However, by combining data from multiple pixels, other parameters can be incorporated into the $Q$-function such as Dark Signal Non-Uniformity (DSNU) and Photo-Response Non-Uniformity (PRNU); expanding the measurement capabilities of the method. Furthermore, assumptions like Poissonian photon shot noise can be relaxed opening up the possibility for PCH-EM algorithms utilizing more exotic photon number distributions. Other opportunities for extending the reliability of the algorithm also exist.  It is understood that PCH-EM is a local optimization strategy prone to converging to a non-optimal local maxima of the likelihood function if provided with poor starting points. As such, future investigations will look into not only expanding the measurement capabilities of the PCH-EM approach, but also relaxing the algorithm's dependence on starting point location; thus, providing a more comprehensive methodology to sensor characterization.


\appendices


\section{Sufficient Statistics}
\label{sec:sufficient_statistics}

Sufficient statistics play a central role in this correspondence and subsequently deserve a quick treatment for the unfamiliar reader. Suppose $\mathbf X=(X_1,\dots,X_N)$ with $X_n\sim F_\theta$ denotes a random sample drawn from a distribution $F_\theta$ parameterized by $\theta$. If $I_\mathbf{X}(\theta)$ denotes the \emph{information} about the unknown parameter $\theta$, contained in $\mathbf X$, and $T(\mathbf X)$ denotes a statistic of $\mathbf X$, e.g., $T(\mathbf X)=\bar X$ (the sample mean), then
\begin{equation}
    I_{T(\mathbf{X})}(\theta)\leq I_\mathbf{X}(\theta)
\end{equation}
with equality only when $T(\mathbf X)$ is a sufficient statistic.  In plain words, if $T(\mathbf X)$ is sufficient, then it captures all of the information about $\theta$ contained in the original sample $\mathbf X$ without loss.

Not all sufficient statistics are created equal.  For example, the original sample $\mathbf X$ is always sufficient but it does not achieve any compression of the data.  For this reason, the category of minimal sufficient statistics are (informally) defined to be sufficient statistics which most efficiently compress the sample $\mathbf X$. In the case of the normal model $X_n\sim\mathcal N(\mu,\sigma^2)$, the minimal sufficient statistic is given by the sample mean and sample variance $T(\mathbf X)=(\bar X,\hat X)$; a two-dimensional statistic. So no matter how large the sample size $N$ is, if $\mathbf X$ is distributed according to a normal distribution, it can always be compressed to a mean and variance (and no further) without incurring any loss of information about $\theta=(\mu,\sigma^2)$; hence, the minimal sufficient statistic $T(\mathbf X)=(\bar X,\hat X)$ can be considered an optimal lossless compression of the sample $\mathbf X$ in the context of parameter estimation.

Working with minimal sufficient statistics has important consequences in deriving quality estimators. Suppose $U$ is an unbiased estimator of a parameter $\theta$ and one defines the estimators $U_1=\mathsf E(U|T_1)$ and $U_2=\mathsf E(U|T_2)$, where $T_1$ is sufficient and $T_2$ is minimal sufficient. The conditional expectations defining $U_1$ and $U_2$ imply these estimators are now functions of $T_1$ and $T_2$, respectively. Then,
\begin{equation}
    \mathsf{Var}U_2\leq \mathsf{Var}U_1
\end{equation}
showing that estimators which are functions of minimal sufficient statistics have less variance than estimators which are functions of ordinary sufficient statistics \cite[Ex.~6.36]{casella_2002}. Indeed, had $T_2$ been complete minimal sufficient, as is the case for the statistic $T(\mathbf X)=(\bar X,\hat X)$ in the normal model, the estimator $U_2$ would represent the unique minimum variance estimator of $\theta$; guaranteeing it incurs less variance than any other unbiased estimator of $\theta$.


\section{Derivation of Constant Flux Multi-Sample PCH-EM Update Equations}
\label{sec:appendix_PCHEM_derivation}

Here, the constant flux multi-sample PCH-EM update equations are derived. Begin with the expected complete data log-likelihood
\begin{multline}
    Q(\theta|\theta^{(t)})=
    \sum_{j=1}^J\sum_{n=1}^{N_j}\sum_{k=0}^\infty \gamma_{jnk}^{(t)}\biggl(-\Phi\tau_j+k\log \Phi\tau_j\\
    -\frac{1}{2}\log s^2-\frac{(x_{jn}-\mu-k/g)^2}{2s^2}+C\biggr),
\end{multline}
where $s=\sigma/g$ and $C$ is a constant independent of $\theta$. The parameterization in terms of $s$ is done for convenience and will simplify the derivation. Before proceeding note that the $\gamma_{jnk}^{(t)}$ are probabilities in $k$ and therefore satisfy
\begin{equation}
\sum_{k=0}^\infty\gamma_{jnk}^{(t)}=1.
\end{equation}

Evaluating $\partial Q/\partial \Phi=0$ and simplifying yields
\begin{equation}
    \frac{1}{N}\sum_{j=1}^J\sum_{n=1}^{N_j}\sum_{k=0}^\infty\gamma_{jnk}^{(t)}(\Phi\tau_j-k)=0.
\end{equation}
Solving this expression for $\Phi$ and recalling the definitions of $\bar{k}^{(t)}$ and $\bar\tau$, the desired form $\Phi^{(t+1)}=\bar{k}^{(t)}/\bar\tau$ is obtained.

Next, evaluating $\partial Q/\partial\mu=0$, dividing both sides by $N$, and simplifying gives
\begin{equation}
    \frac{1}{N}\sum_{j=1}^J\sum_{n=1}^{N_j}\sum_{k=0}^\infty\gamma_{jnk}^{(t)}(x_{jn}-\mu-k/g)=0.
\end{equation}
Upon expanding, further simplification yields
\begin{equation}
    \bar{x}-\mu-\bar{k}^{(t)}/g=0.
\end{equation}
Solving for $\mu$ and using the definition of $\bar{k}^{(t)}$ then yields the update equation $\mu^{(t+1)}$ in terms of $g^{(t+1)}$.

To find the update equation for $g$ evaluate $\partial Q/\partial g=0$, which after some simplification yields
\begin{equation}
    \frac{1}{N}\sum_{j=1}^J\sum_{n=1}^{N_j}\sum_{k=0}^\infty\gamma_{jnk}^{(t)}(x_{jn}-\mu-k/g)k=0.
\end{equation}
Substituting $\mu=\bar{x}-\bar{k}^{(t)}/g$ gives
\begin{equation}
    \frac{1}{N}\sum_{j=1}^J\sum_{n=1}^{N_j}\sum_{k=0}^\infty\gamma_{jnk}^{(t)}(x_{jn}k-\bar{x}k-(k^2-\bar{k}^{(t)}k)/g)=0.
\end{equation}
The trick now is to recognize the identities
\begin{equation}
    \begin{aligned}
        x_{jn}k-\bar{x}k &=(x_{jn}-\bar{x})(k-\bar{k}^{(t)})+\bar{k}^{(t)}x_{jn}-\bar{x}\bar{k}^{(t)}\\
        k^2-\bar{k}^{(t)}k &=(k-\bar{k}^{(t)})^2+\bar{k}^{(t)}k-(\bar{k}^{(t)})^2.
    \end{aligned}
\end{equation}
Since
\begin{equation}
    \frac{1}{N}\sum_{j=1}^J\sum_{n=1}^{N_j}\sum_{k=0}^\infty\gamma_{jnk}^{(t)}(\bar{k}^{(t)}x_{jn}-\bar{x}\bar{k}^{(t)})=0
\end{equation}
and
\begin{equation}
    \frac{1}{N}\sum_{j=1}^J\sum_{n=1}^{N_j}\sum_{k=0}^\infty\gamma_{jnk}^{(t)}(\bar{k}^{(t)}k-(\bar{k}^{(t)})^2)=0
\end{equation}
the equation in question reduces to
\begin{equation}
    \widehat{xk}^{(t)}-\hat{k}^{(t)}/g=0,
\end{equation}
which upon solving for $g$ yields the update equation $g^{(t+1)}$.

To solve the update equation for $\sigma$, it will make the process simpler to first find the update for $s^2$. Evaluating $\partial Q/\partial s=0$ yields after some algebraic manipulations
\begin{equation}
    \frac{1}{N}\sum_{j=1}^J\sum_{n=1}^{N_j}\sum_{k=0}^\infty\gamma_{jnk}^{(t)}(s^2-(x_{jn}-\mu-k/g)^2)=0.
\end{equation}
Substituting $\mu=\bar{x}-\bar{k}^{(t)}/g$ and solving for $s^2$ gives
\begin{equation}
    s^2=\frac{1}{N}\sum_{j=1}^J\sum_{n=1}^{N_j}\sum_{k=0}^\infty\gamma_{jnk}^{(t)}((x_{jn}-\bar{x})-(k-\bar{k}^{(t)})/g)^2.
\end{equation}
Expanding the binomial term yields
\begin{equation}
    s^2=\hat{x}-2\widehat{xk}^{(t)}/g+\hat{k}^{(t)}/g^2.
\end{equation}
Finally, recalling the update equation for $g$ permits writing $\widehat{xk}^{(t)}/g=\widehat{k}^{(t)}/g^2$ so that the update equation for $s^2$ becomes
\begin{equation}
    (s^{(t+1)})^2=\hat{x}-\hat{k}^{(t)}/(g^{(t+1)})^2.
\end{equation}
By the invariance of MLEs and the relation $\sigma=g\sqrt{s^2}$, the update equation for $\sigma$ is obtained.


\section*{Acknowledgment}

Aaron Hendrickson thanks the Naval Innovative Science \& Engineering (NISE) executive board for funding support under project no.~219BAR-24-043.


\bibliographystyle{IEEEtran}
\bibliography{sources}

\end{document}